\documentstyle[11pt,emlines]{article}
\begin{document}

\title{\large \bf GRASERS BASED ON PARTICLE ACCELERATORS AND ON
LASERS} \author{\normalsize E.G.BESSONOV \\ \small \it Lebedev
Physical Institute AS, 117924, Leninsky prospect 53,\\ \small \it
Moscow, Russia \\\small \it E-mail:  bessonov@sgi.lpi.msk.su} \date{}
\maketitle

                           \tableofcontents
\begin{abstract}Grasers based on a stimulated emission of gravitational
radiation by relativistic charged particle beams in external fields and
on conversion of laser radiation into gravitational one in the magnetic
fields as well as detectors are discussed. A scheme of the gravitational
radiation not accompanied by an useless inaccessible by a
value average power of the electromagnetic radiation and stimulation
of the conversion of gravitons into photons in gravitational detectors
by an open resonator are considered.  \end{abstract}

\footskip 10mm
                     \section{Introduction}
The contribution of the gravitational interaction to other kinds of
interactions between elementary particles is very small. That is why to
investigate the nature of the gravitational field and it's interaction
with another fields (conversion of gravitons into photons and other
particles and so on) we are forced to use pure gravitational radiation
for this purpose. On the way to this goal first of all we must prove
the real existence of gravitational waves and study theirs properties.
Then we could solve the problem of the choice of a proper relativistic
theory of gravity, to continue the development of this theory and to
verify it's predictions through new experiments \cite{brill}. If it
will be possible one day the next step could be the investigation of
the significance of the gravity in the elementary particle physics.

Nowadays the efforts on experimental investigation of foundations of
the theory of gravity are concentrated on the detection of
gravitational radiation coming from the universe. However the
wavelengths of the existing natural sources of the gravitational
radiation like double stars is very high (period is about 1 day). The
intensity of these and other more hard natural gravitational radiation
sources is very small \cite{bragin}. Phenomena similar to Supernovae
explosions (when much higher power and frequency ($\sim 1 kHz$)
gravitational radiation is emitted) don't often happen. That is why we
are forced to produce an artificial source of hard gravitational
radiation. One of the possible versions of such source is the
gravitational analogue of Laser named a Graser. The necessity in
Grasers is similar to that in nuclear research when particle
accelerators began to be used instead of cosmic rays.

In this paper both Grasers based on a direct emission of
gravitational waves by nonuniformly moving prebunched ion beams in
undulators (parametric Free-Ion Grasers) and Grasers based on the
conversion of gaussian laser beams stored in an open resonator into
gravitational beams in a transverse magnetostatic fields are
considered. It was shown that one of optimal solutions of the problem
of a Graser based on direct emission of gravitational radiation by
particles can be a Graser based on an undulator, a cutoff waveguide
arranged in the undulator and a relativistic heavy ion beam.  Such
Graser is working under conditions when a gravitational radiation is
not accompanied by the useless electromagnetic radiation with an
inaccessible by a value average power. In the relativistic case these
conditions are valid at wavelengths much less then the cutoff one.
An effective method of stimulation of the conversion process of
gravitons into photons in a transverse magnetic field, multilayer
mirrors  and other systems of gravitational detectors by a resonator
are considered.

          \section{Emission of gravitational radiation}
Weak gravitational processes in a weak gravitational field of the Earth
can be described in the Euclidian space. In static the Coulomb's law of
forces between electric charges and Newton's law of the universal
gravitation have the same dependence of forces on a distance $R$
between particles ($\sim 1/R^2$) and on theirs charges $\; $e$\;$ and
masses $\; M$. It means that in the case of gravitational field we
can introduce the gravitational charge $e_{gr} = \sqrt{G}\, M$ and use
it in the static gravity theory the same way as the electric charge is
used in the electrostatic, where $G = 6.67\,10^{-8}$ is the
gravitational constant. The difference in the formulas of these lows is
only in the sign. Gravitational field is always attractive field for
all particles.

We can extend this analogy from static to dynamic of charges, forces
and fields and suppose that gravitational field of moving particle in
general case (that is in an arbitrary reference frame and arbitrary
particle motion) will have both the analogy of the electric $\vec
E^{gr}$ and analogy of the magnetic $\vec B^{gr}$ field strengths
determined by a gravitational charge $e_{gr}$, velocity and
acceleration. If we proceed from this supposition and from the
equivalence of the inertial reference frames then in order to this
analogy was fruitful we must determine the dynamical lows of gravity.
At that we can proceed from the analogy with the dynamical approach to
the construction of the special theory of relativity (G.Lorentz,
A.Poincare) \cite{e.l.f}.  The simplest low can be found by considering
some particular problems with known solution. In particular we know
that a system of 2 identical particles with masses $M$ and in a charge
state $n^{+}$ will be in equilibrium in any arbitrary moving coordinate
system when it is in equilibrium in a coordinate system at rest
($G\,M^2 = (n^{+})^2 e^2$).  If the gravitational force is higher
($G\,M^2 > (n^{+})^2 e^2$) then such system will be in the state of
rotation. At that in accordance with the relativistic transformations
the longitudinal dimension of the orbits of particles will be
compressed and the period of rotation will be increased relative to the
laboratory coordinate system when the center of mass is moving with a
relativistic velocity.  It means that the electromagnetic and
gravitational transformations of forces and fields must be the same and
the equations of gravity  must be similar to the Maxwell's equations
\cite{mitskevich}, \cite{petrov}.  Otherwise the laws of the particle
motion will have different forms in different rest frames moving with
different velocities if gravitational, electromagnetic and any other
forces between particles will have different transformation properties.

So in a general case time-varying week gravitational analogies of
electric and magnetic fields in the first approximation are described
by the wave equations which are similar to the wave equations of the
electromagnetic fields (with the accuracy to signs of terms of the
equations which include the density of charges). The solution of the
gravitational equations for the nonuniformly moving particle which
undergo the influence of the extraneous fields will lead to the
gravitational analogies of the electromagnetic fields of the form
similar to Lienard-Viechert's form for the electromagnetic fields.

It follows that the nonuniformly moving charged particle will emit the
gravitational radiation with the same parameters (except power) as the
electromagnetic radiation of these particles in free space (the
influence of the boundary conditions for the gravitational radiation is
negligible). That is why we can declare that Free-Particle Lasers will
emit stimulated gravitational radiation simultaneously with the
stimulated electromagnetic radiation. These kinds of radiation
emitted in the conditions of free space will have identical spectral and
angular distributions\footnote{According to this concept we can draw
the conclusion that all sources of the electromagnetic radiation
(thermal sources, lasers, klystrons and so on) are the sources of the
gravitational radiation simultaneously. The difference between emission
of electrons in undulator of Free-Electron Laser and in an atom is in
the fact that a nucleon emit gravitational wave of the same amplitude
as an electron but opposite polarity (theirs accelerations have
opposite directions and inverse to theirs masses and the emitted field
strengths are proportional to theirs masses and accelerations). As the
wavelength of the emitted radiation in this case is much higher then
the dimension of the atom then the difference between phases of the
emitted radiation and $\pi$ is negligible and the power of the emitted
radiation will be suppressed.  It will appear in the approximation
$\beta ^2 = (v/c)^2$, where $v$, $c$ are the velocities of the electron
and light accordingly. In this case in the correlation (1) a
coefficient $C_{\alpha} = C_{\alpha 0} \simeq \beta ^2 \ll 1$ will
appear. However in the case of an extended media photons emitted by an
atom can be absorbed by another atoms of media with high probability
and gravitons will go away from media with very small absorption. That
is why for the case of lasers based on high quality resonators and
without extraction of the emitted electromagnetic radiation or large
sources like stars the ratio of the gravitational to electromagnetic
radiation will be much higher ($C_{\alpha} \gg C_{\alpha 0} $).}. The
ratio of the power of the gravitational to electromagnetic radiation
will be the square of the ratio of their's charges:  $P^{gr}/P^{em} =
e_{gr}^2/(n^{+} e)^2 = GM^2/(n^{+}e)^2$.  Different gravitational
theories lead to ratio

                 \begin{equation}       
       {P^{gr}\over P^{em}} = {C_{\gamma}GM^2\over
                 (n^{+})^2e^2}
                 \end{equation}
with the coefficient $C_{\gamma} \simeq 1 \div 3$. For example in the
case of Einstein's theory of gravity $C_{\gamma} \simeq 13/4$
\cite{pust}.  The values $GM^2/e^2 = 2.402\cdot10^{-43}$ for electron
mass $m_e = 9.11 \cdot 10^{-28}$ g and $GM^2/e^2 = 7.98 \cdot
10^{-37}$ for unit atomic mass $M_u = 1.66 \cdot 10^{-24}$ g.

The flow of the monochromatic gravitons of the frequency $\omega _{gr}$

                 \begin{equation}         
        \dot N^{gr} = {P^{gr}\over \hbar \omega _{gr}} = {C_{\gamma} G
        M^2\over (n^{+})^2 e^2\hbar \omega _{gr}} P^{em} = {C_{\gamma}
        G M^2\over (n^{+})^2 e^2}\dot N^{em},
              \end{equation}
where $\hbar $ is the Plank reduced constant, $\dot N^{em} \simeq
5\cdot 10^{22} P^{em}\lambda $\,[ph/W$\cdot$cm] the flow of the laser
photons, $\lambda = 2\pi c/\omega _{gr}$.

          \subsection{\it Grasers based on particle emission in
                    undulators}

The total energy, hardness and directivity of the
gravitational radiation of particles emitted in the magnetic fields of
undulators is strongly increased with the increase of the relativistic
factor $\gamma$. The number of the emitted both photons and gravitons
in this case does not depend on $\gamma$.  Free-Ion Lasers based on
proton and more heavy ion beams of storage rings can emit more powerful
coherent electromagnetic and gravitational radiation then Free-Electron
Lasers of the same relativistic factor \cite{bes1} - \cite{bes-kim}.
This is because of the total energy of ion beams stored in ion storage
rings is much higher then the total energy of electron beams stored in
electron rings. Grasers based on ion beams according to (1) have an
additional advantage as the ratio of the gravitational to
electromagnetic powers of the emitted radiation is proportional to the
square of mass. This is important as a weak power of the gravitational
radiation will be accompanied by the extremely high useless power of
the electromagnetic radiation. The value of the power of the
electromagnetic radiation have the inaccessible high value for
electrons and light ions. Very hard conditions are in the case of heavy
ions up to ${^{238}_{92}U^{1+}}$ and more heavy formations as well. For
example if we accept the power of the electromagnetic radiation emitted
by an ion beam $P^{em} = 10^8$ W, $M = 238 M_u$, $n^{+} =1$,
$C_{\gamma} = 13/4$, ($C_{\gamma}GM_ {^{238}_{92}U^{1+}}^2/e^2 = 1.47
\cdot 10^{-31}$) then the power of the gravitational radiation is
$P^{gr} = 1.47 \cdot 10^{-23}$ W, the flow of gravitons $\dot N^{gr} =
7.35 \cdot 10^{-5}$ gr/s when the energy of the gravitons is $\hbar
\omega _{gr} = 1.25 eV = 2\cdot 10^{-19}$ J ($\lambda_{gr} \sim 1 \mu$)
and $\dot N^{gr} = 0.74$ gr/s when the energy of the gravitons is
$\hbar \omega _{gr} = 1.25 \cdot 10^{-4} eV = 2\cdot 10^{-23}$ J
($\lambda_{gr} \sim 1 cm$). The production of the same flow of
gravitons in the \,$\lambda \leq 1$ cm, region by Grasers based on
Free-Electron Lasers ($C_{\gamma}Gm_e^2/e^2 = 7.81 \cdot 10^{-43}$)
require an expense of average power $1.88\cdot 10^{11}$ times higher.
This is not practicable solution.

Production of the flow of the gravitational radiation $\sim 1$ gr/s is
not enough to detect them now. To produce higher flows we need the way
of the gravitational beam production which is not accompanied by the
emission of the electromagnetic radiation of the inaccessible by the
value average power. One of possible solutions of this problem is in an
using of the cutoff waveguides (regime of cutoff or evanescent modes
\cite{jackson}). In this case only the gravitational radiation will be
emitted (it does not interact with waveguide).

A scheme of a Graser and at the same time a scheme of a Parametric
(Prebunched) Free-Ion Laser based on an undulator and prebunched Ion
Beam of a storage ring is presented on the Fig.1. Parametric Free-Ion
Lasers have an advantage on ordinary Free-Ion Lasers originating from
the prebunched nature of ion beams\footnote{In ordinary Free-Ion Lasers
rather long distance is necessary before a homogeneous ion beam will be
bunched when starting from noise (self amplified stimulated emission)
or week external electromagnetic wave.  Only after bunching the ion
beam can effectively produce radiation.  Moreover the degree of the
beam bunching in these cases is not high.  Because of this factors the
Parametric Free-Particle Lasers based on high degree bunched beams have
limiting parameters for the rate of the energy loss.  They will emit
monochromatic, polarized, diffractionally limited radiation as well
\cite{bes1} - \cite{bes-kim}, \cite{bes4},
\cite{bes5}.}.
\vspace{9mm}
\begin{center}
\special{em:linewidth 0.4pt}
\unitlength 0.90mm
\linethickness{0.4pt}
\begin{picture}(160.67,112.33)
\put(148.00,100.00){\vector(1,0){1.12}}
\emline{10.00}{100.00}{1}{148.00}{100.00}{2}
\emline{24.67}{105.00}{3}{22.50}{104.91}{4}
\emline{22.50}{104.91}{5}{20.51}{104.63}{6}
\emline{20.51}{104.63}{7}{18.68}{104.18}{8}
\emline{18.68}{104.18}{9}{17.02}{103.54}{10}
\emline{17.02}{103.54}{11}{15.54}{102.72}{12}
\emline{15.54}{102.72}{13}{14.22}{101.71}{14}
\emline{14.22}{101.71}{15}{13.08}{100.53}{16}
\emline{13.08}{100.53}{17}{12.10}{99.16}{18}
\emline{12.10}{99.16}{19}{11.30}{97.60}{20}
\emline{11.30}{97.60}{21}{10.67}{95.87}{22}
\emline{10.67}{95.87}{23}{10.20}{93.95}{24}
\emline{10.20}{93.95}{25}{9.91}{91.85}{26}
\emline{9.91}{91.85}{27}{9.79}{89.57}{28}
\emline{9.79}{89.57}{29}{10.00}{85.00}{30}
\emline{25.00}{95.00}{31}{23.19}{94.44}{32}
\emline{23.19}{94.44}{33}{21.78}{93.44}{34}
\emline{21.78}{93.44}{35}{20.75}{92.00}{36}
\emline{20.75}{92.00}{37}{20.11}{90.11}{38}
\emline{20.11}{90.11}{39}{19.86}{87.78}{40}
\emline{19.86}{87.78}{41}{20.00}{85.00}{42}
\emline{25.00}{105.00}{43}{25.00}{95.00}{44}
\emline{20.00}{85.00}{45}{10.00}{85.00}{46}
\emline{115.00}{105.00}{47}{115.00}{95.00}{48}
\emline{115.00}{95.00}{49}{116.90}{94.08}{50}
\emline{116.90}{94.08}{51}{118.37}{92.85}{52}
\emline{118.37}{92.85}{53}{119.42}{91.34}{54}
\emline{119.42}{91.34}{55}{120.04}{89.52}{56}
\emline{120.04}{89.52}{57}{120.23}{87.41}{58}
\emline{120.23}{87.41}{59}{120.00}{85.00}{60}
\emline{115.00}{105.00}{61}{117.20}{104.81}{62}
\emline{117.20}{104.81}{63}{119.22}{104.43}{64}
\emline{119.22}{104.43}{65}{121.08}{103.86}{66}
\emline{121.08}{103.86}{67}{122.76}{103.09}{68}
\emline{122.76}{103.09}{69}{124.26}{102.14}{70}
\emline{124.26}{102.14}{71}{125.59}{101.00}{72}
\emline{125.59}{101.00}{73}{126.75}{99.67}{74}
\emline{126.75}{99.67}{75}{127.73}{98.14}{76}
\emline{127.73}{98.14}{77}{128.55}{96.43}{78}
\emline{128.55}{96.43}{79}{129.18}{94.52}{80}
\emline{129.18}{94.52}{81}{129.65}{92.43}{82}
\emline{129.65}{92.43}{83}{129.94}{90.14}{84}
\emline{129.94}{90.14}{85}{130.06}{87.67}{86}
\emline{130.06}{87.67}{87}{130.00}{85.00}{88}
\emline{32.67}{97.00}{89}{25.00}{97.00}{90}
\emline{25.00}{97.00}{91}{22.98}{96.67}{92}
\emline{22.98}{96.67}{93}{21.27}{96.00}{94}
\emline{21.27}{96.00}{95}{19.86}{95.00}{96}
\emline{19.86}{95.00}{97}{18.75}{93.67}{98}
\emline{18.75}{93.67}{99}{17.94}{92.00}{100}
\emline{17.94}{92.00}{101}{17.43}{90.00}{102}
\emline{17.43}{90.00}{103}{17.23}{87.67}{104}
\emline{17.23}{87.67}{105}{17.33}{85.00}{106}
\emline{17.33}{66.67}{107}{17.33}{85.00}{108}
\put(69.67,110.00){\makebox(0,0)[cc]{{\small $Undulator$} }}
\put(13.67,86.99){\makebox(0,0)[cc]{{\small $BM$}}}
\put(126.67,87.00){\makebox(0,0)[cc]{{\small $BM$}}}
\put(121.63,80.50){\oval(2.00,1.00)[]}
\emline{123.00}{85.00}{109}{123.00}{66.33}{110}
\put(121.63,75.50){\oval(2.00,1.00)[]}
\put(121.63,70.50){\oval(2.00,1.00)[]}
\put(134.00,75.33){\makebox(0,0)[cc]{\small $ion$ $bunches$}}
\put(16.00,80.50){\oval(2.00,1.00)[]}
\put(16.00,75.50){\oval(2.00,1.00)[]}
\put(16.00,70.50){\oval(2.00,1.00)[]}
\put(28.67,75.33){\makebox(0,0)[cc]{\small $ion$ $bunches$}}
\put(75.00,50.33){\makebox(0,0)[cc]
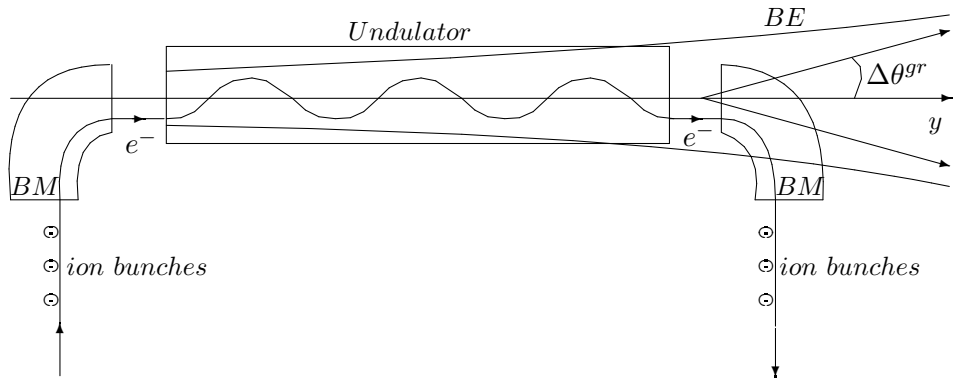
{{\small Fig.1: A scheme of the parametric Free-Ion Laser; BM: Banding Magnets; BE: Beam}}}
\put(75.00,45.33){\makebox(0,0)[cc]
{{\small Envelope of the electromagnetic and gravitational radiation; $e^{-}$ electron bunches.}}}
\put(148.67,110.00){\vector(4,1){0.2}}
\emline{112.00}{100.00}{111}{148.67}{110.00}{112}
\put(148.67,90.00){\vector(4,-1){0.2}}
\emline{112.33}{100.00}{113}{148.67}{90.00}{114}
\emline{134.33}{106.00}{115}{135.51}{104.70}{116}
\emline{135.51}{104.70}{117}{135.71}{102.87}{118}
\emline{135.71}{102.87}{119}{134.67}{100.00}{120}
\put(141.00,103.67){\makebox(0,0)[cc]{$\Delta \theta ^{gr} $}}
\put(146.67,96.00){\makebox(0,0)[cc]{{\small $y$}}}
\put(123.00,59.00){\vector(0,-1){0.2}}
\emline{123.00}{66.00}{121}{123.00}{59.00}{122}
\put(17.33,66.67){\vector(0,1){0.2}}
\emline{17.33}{59.00}{123}{17.33}{66.67}{124}
\put(124.33,112.00){\makebox(0,0)[cc]{{\small $BE$}}}
\emline{33.00}{93.33}{125}{107.33}{93.33}{126}
\emline{107.33}{93.33}{127}{107.33}{107.67}{128}
\emline{107.33}{107.67}{129}{33.00}{107.67}{130}
\emline{33.00}{107.67}{131}{33.00}{93.33}{132}
\put(29.34,93.66){\makebox(0,0)[cc]{$e^{-}$}}
\emline{107.33}{97.00}{133}{115.00}{97.00}{134}
\emline{120.00}{85.00}{135}{130.00}{85.00}{136}
\emline{123.00}{85.00}{137}{122.88}{87.81}{138}
\emline{122.88}{87.81}{139}{122.50}{90.25}{140}
\emline{122.50}{90.25}{141}{121.88}{92.31}{142}
\emline{121.88}{92.31}{143}{121.00}{94.00}{144}
\emline{121.00}{94.00}{145}{119.88}{95.31}{146}
\emline{119.88}{95.31}{147}{118.50}{96.25}{148}
\emline{118.50}{96.25}{149}{116.88}{96.81}{150}
\emline{116.88}{96.81}{151}{115.00}{97.00}{152}
\emline{33.00}{97.00}{153}{35.20}{97.15}{154}
\emline{35.20}{97.15}{155}{37.19}{98.02}{156}
\emline{37.19}{98.02}{157}{39.33}{100.00}{158}
\emline{39.33}{100.00}{159}{40.95}{101.58}{160}
\emline{40.95}{101.58}{161}{42.90}{102.58}{162}
\emline{42.90}{102.58}{163}{45.67}{103.00}{164}
\emline{45.67}{103.00}{165}{47.81}{102.62}{166}
\emline{47.81}{102.62}{167}{51.67}{100.00}{168}
\emline{51.67}{100.00}{169}{53.58}{98.28}{170}
\emline{53.58}{98.28}{171}{55.55}{97.27}{172}
\emline{55.55}{97.27}{173}{58.00}{97.00}{174}
\emline{58.00}{97.00}{175}{60.20}{97.15}{176}
\emline{60.20}{97.15}{177}{62.19}{98.02}{178}
\emline{62.19}{98.02}{179}{64.33}{100.00}{180}
\emline{83.00}{97.00}{181}{85.20}{97.15}{182}
\emline{85.20}{97.15}{183}{87.19}{98.02}{184}
\emline{87.19}{98.02}{185}{89.33}{100.00}{186}
\emline{64.33}{100.00}{187}{65.95}{101.58}{188}
\emline{65.95}{101.58}{189}{67.90}{102.58}{190}
\emline{67.90}{102.58}{191}{70.67}{103.00}{192}
\emline{89.33}{100.00}{193}{90.95}{101.58}{194}
\emline{90.95}{101.58}{195}{92.90}{102.58}{196}
\emline{92.90}{102.58}{197}{95.67}{103.00}{198}
\emline{70.67}{103.00}{199}{72.81}{102.62}{200}
\emline{72.81}{102.62}{201}{76.67}{100.00}{202}
\emline{95.67}{103.00}{203}{97.81}{102.62}{204}
\emline{97.81}{102.62}{205}{101.67}{100.00}{206}
\emline{76.67}{100.00}{207}{78.58}{98.28}{208}
\emline{78.58}{98.28}{209}{80.55}{97.27}{210}
\emline{80.55}{97.27}{211}{83.00}{97.00}{212}
\emline{101.67}{100.00}{213}{103.58}{98.28}{214}
\emline{103.58}{98.28}{215}{105.55}{97.27}{216}
\emline{105.55}{97.27}{217}{108.00}{97.00}{218}
\put(29.67,97.00){\vector(1,0){0.2}}
\emline{28.00}{97.00}{219}{29.67}{97.00}{220}
\put(112.00,97.00){\vector(1,0){0.2}}
\emline{110.33}{97.00}{221}{112.00}{97.00}{222}
\emline{33.00}{104.00}{223}{74.96}{105.98}{224}
\emline{74.96}{105.98}{225}{103.11}{107.80}{226}
\emline{103.11}{107.80}{227}{120.81}{109.26}{228}
\emline{120.81}{109.26}{229}{133.53}{110.51}{230}
\emline{133.53}{110.51}{231}{142.23}{111.50}{232}
\emline{142.23}{111.50}{233}{148.67}{112.33}{234}
\emline{148.67}{87.00}{235}{146.04}{87.52}{236}
\emline{146.04}{87.52}{237}{143.33}{88.02}{238}
\emline{143.33}{88.02}{239}{140.54}{88.51}{240}
\emline{140.54}{88.51}{241}{134.73}{89.44}{242}
\emline{134.73}{89.44}{243}{128.59}{90.31}{244}
\emline{128.59}{90.31}{245}{122.13}{91.12}{246}
\emline{122.13}{91.12}{247}{115.35}{91.87}{248}
\emline{115.35}{91.87}{249}{104.59}{92.88}{250}
\emline{104.59}{92.88}{251}{93.10}{93.76}{252}
\emline{93.10}{93.76}{253}{76.66}{94.71}{254}
\emline{76.66}{94.71}{255}{54.31}{95.56}{256}
\emline{54.31}{95.56}{257}{33.00}{96.00}{258}
\put(111.68,94.66){\makebox(0,0)[cc]{$e^{-}$}}
\end{picture}
             \end{center}
            \vspace{-34mm}
Below we will consider the possible parameters of such Grasers. First
of all we will present some consequences from the theory of Lasers and
Grasers.

The minimum wavelength $\lambda_n$ of the electromagnetic radiation
emitted by a relativistic charged particle in an undulator on the
$n$-th harmonic in free space and in a waveguide is defined by an
undulator period $\lambda_u$, relativistic factor $\gamma $, deflecting
parameter $p_\perp$ and the cutoff wavelength of the waveguide mode
$\lambda _c$:

          $$\lambda_n^{fs}={\lambda_u\over2n\gamma^2}(1+p_\perp^2), $$
                        \begin{equation}  
          \lambda_n^{wg} = \frac {\lambda_u \left (1\pm\sqrt {1-(1+p_\perp
            ^2)(1+{\lambda _u^2/n^2\lambda _c^2})/\gamma^2}\right
            )}{\overline{\beta_y}n\left(1 +
          {\lambda _u^2/n^2\lambda _c^2}\right )},
           \end{equation}  
where $\overline{\beta_y} = \overline{(\beta ^2 - \beta_{\perp}^2)
^{1/2}}$, $p_\perp = \left(\overline {|\vec p_\perp |^2}\right)^{1/2} =
\left(\overline {|\vec B_{\perp }|^2}\right )^{1/2} /B_c$, $\vec
p_\perp $ $=\gamma \vec \beta_ \perp,$ $ \vec \beta_\perp=\vec
v_\perp/c$, $\vec v_\perp$ is a transverse velocity of the particle,
$\vec B_{\perp}$ a transverse magnetic field strength of an undulator,
$B_c = 2\pi Mc^2/n^+ e\lambda _u$,\, $e$\, an electron charge, $n^+$ a
number of an ion charge state, a coefficient $2\pi M c^2/e\simeq 19.56
(M/M_u)[MG\cdot cm]$ \cite{al-bes}.

According to (3) the electromagnetic radiation will be emitted in the
waveguide when the relativistic factor of particles $\gamma >
\sqrt{(1+p_{\perp}^2)(1+\lambda_u^2/\lambda_c^2)}$.  In this case the
wavelengths of the emitted radiation $\lambda < \lambda _{max}^{wg}$
where

                        \begin{equation}  
          \lambda_{max}^{wg}|_{n=1} = \frac {\lambda_u \lambda _c^2}
          {\lambda _u^2 + \lambda _c^2}|_{\lambda _u \gg \lambda _c}
          \simeq {\lambda _c^2\over \lambda _u} \simeq {\lambda _c
          \sqrt{1+p_{\perp}^2}\over\gamma}.  \end{equation} 

The emitted radiation is diffraction limited. The angular divergence of
the gravitational beam $\Delta \theta ^{gr}$ is determined by the
relativistic factor $\gamma$, the emitted wavelength $\lambda $, the
particle beam radius $\sigma _{\perp}$ and the number of the undulator
periods $K$:

         \begin{equation} 
       \Delta \theta ^{gr} = min({\lambda \over \sigma _{\perp}},
       {1 \over \gamma }\sqrt {1+ p_{\perp}^2\over nK}).  \end{equation}

The power emitted by an ion beam on the first harmonic in a Parametric
Free-Ion Laser based on a helical undulator ($\left(\overline {|\vec
B_{\perp }|^2}\right )^{1/2} = B_{\perp}$) in the regime of free space
when using a particle beam consisting of a series of short ($<\lambda _1$)
microbunches with a small transverse dimensions $\sigma _p \ll
\lambda \gamma/\sqrt {1+p_\perp ^2}$, spaced an integral number of
wavelengths $\lambda _1$, apart is,

         \begin{equation} 
       P^{em\,fs} = {\pi ^2 \over c}{p_{\perp}^2\over
       1 + p_{\perp}^2} K i^2   \qquad \mbox{or} \qquad P^{em\,fs} =
       296.1 {p_{\perp}^2 \over 1 + p_{\perp}^2} K i^2 [{W\over A^2}],
         \end{equation}
where $i$ is the electron beam current \cite{bes2}, \cite{bes3}.
When the wide beam of the radius $\sigma _p \gg \lambda \gamma/\sqrt
{1+p_\perp ^2}$ is used then the power tend to (6) at the distances
$l_p = 2\pi \sigma _p^2/\lambda _1$ from the beginning of the
undulator.

Accordingly, the power of the gravitational radiation and the flow of
gravitons in the case of $C_{\gamma} = 13/4$ are:

         \begin{equation} 
       P^{gr} = 7.68\cdot 10^{-34}{p_{\perp}^2 \over 1 +
       p_{\perp}^2}({M\over M_u})^2{K i^2\over (n^{+})^2},
       \quad \dot N^{gr} = 4.79\cdot 10^{-15} {p_{\perp}^2 \over
       1 + p_{\perp}^2}({M\over M_u})^2{K i^2\over (n^{+})^2 \hbar
       \omega},  \end{equation}
where $P$ is in Watts, $i$ in Amperes and $\hbar \omega $ in eV
\cite{bes2}, \cite{bes3}.

Notice that in the relativistic case the power of the emitted coherent
radiation (7) does not depend on the energy of particles ($\gamma $)
and the flow of gravitons $\dot N^{gr} \sim \gamma ^{-2}$.

The value $\lambda _{max}|_{\lambda _u \gg \lambda _c} \ll \lambda _c$.
It means that we can generate hard gravitational radiation $\lambda \ll
\lambda _c$ under conditions when it does not accompany the
electromagnetic one. The emission of higher harmonics in this Graser
must be absent as well. Undulator with an ideal helical magnetic field
will satisfy this condition when particle bunches emit radiation in
phase in the direction of the axis. \vskip 3mm

            \vskip 5mm {\it \bf Example 1.}\hskip 4mm
The beam of $^{238}_{92}U^{1+}$ ions pass through the waveguide which
has the form of a smooth pipe of the radius $r=1.5$ cm. The waveguide
is installed into a helical undulator with a period $\lambda_u$ = 46.55
m, a number of periods $16$ (a length 745 m) and magnetic field
strength $B_\perp$ = 300 k\,G. The energy of ions in the storage
ring is $\varepsilon _i \simeq 11.1$ TeV ($\gamma$ = 50.4) and the
current of the ion beam is $i$ = 1 kA.

In this case the fundamental $H_{11}$ mode has a cutoff wavelength
$\lambda _c$ $\simeq $ 3.41 $r$ $\simeq $ 5.1 cm, $\lambda _{max}^{wg}
\simeq 5.6 \cdot 10^{-3}$ cm, $B_c \simeq 10^6$ G, $p_\perp \simeq
0.3,$  $n$ = 1. According to (3), (5), (6) the wavelength of the
coherent radiation is $\lambda _1 \simeq $ 1 cm, the energy of a
graviton $1.25\cdot 10^{-4}$ eV, the angular divergence of the
gravitational beam is $\Delta \theta \simeq 5\cdot 10^{-3}$, the power
of the gravitational radiation and the flow of the gravitons emitted by
the ion beam are:  $P^{gr} = 5.7 \cdot 10^{-23}$ W, $\dot N^{gr} =
2.92$ gr/s.

If the energy of ions in the storage ring is $\varepsilon _i \simeq
57.5$ TeV ($\gamma$ = 260) then the wavelength of the coherent
radiation is $\lambda _1 \simeq $ 3.76$\cdot 10^{-2}$ cm, the energy of
a graviton $3.32\cdot 10^{-3}$ eV, the power of the gravitational
radiation and the flow of the gravitons emitted by the ion beam are:
$P^{gr} = 2.4 \cdot 10^{-23}$ W, $\dot N^{gr} = 0.11$\,gr/s,  $\Delta
\theta \simeq \cdot 10^{-3}$.

In these two cases $\lambda > \lambda _{max}^{wg}$ the unwanted power
of the electromagnetic radiation ($P^{em} = 46.9$\,MW) will not be
emitted.

               \vskip 5mm
Spectra of both electromagnetic and gravitational radiation sources are
limited from the short wavelength region by the bunch length. In
modern storage rings electron bunches of 0.3 mm (1-ps) length were
achieved.  The longer-term objective is the realization of 100-fs
bunches \cite{krinsky} - \cite{nim92}. In linear accelerators using
electron guns with both thermal and photoelectron cathodes and bunchers
the production of electron bunches of the length $\sim 0.1\div 0.01 $
mm was achieved \cite{wiedemann} - \cite{serafini}. Bunchers based on
undulators and lasers permit to produce a long train of more short
microbunches. The bunching frequency in this case is very high (the
distance between bunches is equal or multiple to the wavelength of the
laser beam). To prepare ion beams for Grasers in ion storage rings the
same beam bunching problems can be solved by the same or special
bunching technique \cite{microb}. Instead of a three dimensional
Synchrotron Radiation Damping using in electron storage rings a
Radiative Ion Cooling based on the process of resonance Rayleigh
scattering of a laser light by relativistic ions \cite{prl96},
\cite{bes3}, \cite{bes-kim} or ordinary Laser Cooling under conditions
of synchro-betatron resonance \cite{sessler}, \cite{bes6}, \cite{bes7}
can be used in ion storage rings. Ion cooling may make it possible to
store very high current low-emittance ion beams using multiple
injection of ions in high energy storage rings. The beam stored energy
of the LHC will exceed the value 500 MJ (average current $\sim 0.5 A$).
Multiple injection at the top energy of the dedicated ion storage ring
will permit to increase this value. The energy stored in the proton
beam at the top energy of the ISR has reached $5\cdot 10^6$ J, in the
form of a beam of 50 A at 31.5 GeV.  The peak currents in excess of
$10^3$ A have been focused to submillimetric spots without problems
\cite{rubbia}.

        \subsection{\it Grasers based on conversion of a laser beam into
        a gravitational one in a transverse magnetostatic field}

In section 2 we introduced gravitational charge $e_{gr}$ for a particle
with a mass $M$. We can do the next step and suppose that all forms of
fields enable a density of energy $\rho = w/c^2$ and a corresponding
charge density $\rho _{gr} = \sqrt{G}\rho$. In the case of
electromagnetic fields $w = (|\vec E|^2 + |\vec H|^2)/8\pi$, where
$\vec E$, $\vec H$ are the electric and magnetic field strengths.  If
the density of the energy of the electromagnetic field is varied with
time then a varying with time gravitational charge density will appear
and the conditions for the emission of the gravitational radiation can
appear. For example, the superposition of static electromagnetic field
and electromagnetic wave in vacuum will lead to a formation of a
modulated gravitational charge density propagating together with
electromagnetic wave the same direction with light velocity and to
emission of gravitational wave.

Electrogravitational conversion of an electromagnetic radiation to
gravitational one was known to Whittaker as early as 1947 \cite{whitt}.
Gertsenshtein was the first to actually calculate the conversion
efficiency of a plane electromagnetic wave into gravitational one in a
transverse magnetic field \cite{gerts}, \cite{logi}. The conversion of
free electromagnetic wave into gravitational one is lacking but in the
simplest case of two counter propagating waves it take place
\cite{grisch1}, \cite{grisch2}.

The way based on the conversion of the electromagnetic laser beam into
the gravitational one in a one-directional transverse magnetic field is
presented on Fig.2. On this figure an open resonator presented by two
mirrors $M_1$, $M_2$. The resonator is located outside of the magnet
poles. The distance between mirrors $L_m$ is higher then the length of
the magnet $L$. It can be exited by continuous beam of Free Electron or
Free-Ion Laser. Conversion of electromagnetic beam into gravitational
one takes place in the magnetic field. The envelop of the gravitational
beam coincides with the mode envelop of the electromagnetic radiation
in the resonator. Outside of the resonator the electromagnetic
radiation is absent but the gravitational radiation is propagated in
the limits of the prolonged resonator mode envelop.

The ratio of the power of the gravitational radiation to the active
power of the electromagnetic radiation stored in the resonator that is
the probability $p _{\gamma \to g}$ of the conversion

                 \begin{equation}       
       p_{\gamma \to g} =  {P^{gr}\over P^{em}_a} = {G\over
             c^4}Q(\int _0^L B_{\perp}(y)dy)^2, \end{equation}
where $G/c^4 \simeq 8.24 \cdot 10^{-50}$[1/cm$^2$\,Gauss$^2$], $Q$
is the quality of the resonator, $L$ the length of the magnet,
$B_{\perp}$ the value of the transverse magnetic field \cite{gerts}.

It follows from the equation (8) that the effective conversion takes
place in the case of single sign transverse component of the magnetic
field $(\int B_{\perp}dy)^2 = \overline{B_{\perp}^2}L^2 $). The
gravitational radiation will not be emitted in the magnetic field of
the undulator because of in the undulator $\int B_{\perp}dy = 0$ (the
waves emitted in the magnetic poles of undulator of different polarity
will have phase shift $\pi $).

The mode distribution at an arbitrary point of resonator is determined
by the expression

         \begin{equation} 
       \sigma = \sigma _0 \sqrt {1 + {y\over l_R}},
        \end{equation}
where $l_R = {\pi \sigma _0^2/\lambda }$ is the Rayleigh length,
$\sigma _0$ the radius of the waist of a photon beam. In the case of
confocal resonator: $L_m = 2l_R$ and $\sigma _0 = \sqrt{\lambda
L_m/2\pi}$. Photon and gravitational beams radii are increased $\sqrt
2$ times and theirs area $2$ times per the length $l_R$. At the
distance $\sigma $ the energy density of beams in the transverse
direction decay \,$e$\, times \cite{maitland}.

The following example illustrates possible parameters of such Graser.

         \vskip 5mm  {\it \bf Example 2.}\hskip 4mm
Let the value of the magnetic field is $B_{\perp} = 10^5$\,G, the
length of the bending magnet $L = 1$\,km, the active losses in the
cavity $P^{em}_a = 10^5$\,W, the quality of the cavity $Q =
10^3$, the reactive power (which determine the conversion process) is
$P^{em}_r$ = 100 \,MW.

In this case according to (8), (9) the power of the gravitational
radiation is $ P^{gr} = 8.26 \cdot 10^{-22}$\,W, the flow of the
gravitons and the radius of the waist of the photon (graviton) beams
are $\dot N^{gr} = 4\cdot 10^{-3}$ gr/s, $\sigma _0 = 1.26$\,cm and
$\dot N^{gr} = 40 $ gr/s, $\sigma _0 = 126$\,cm for the photon energy
1.25 eV and $1.25 \cdot 10^{-4}$\,eV accordingly. The main problem in
this scheme is the heating problem of mirrors. The figures above can be
increased by increasing of the length of the magnet, the value of the
magnetic field if it will be possible in future developments of the
superconducting technique and the quality of resonators in the
centimeter and lower wavelength regions.

             \vspace{25mm}
             \begin{center}
\special{em:linewidth 0.4pt}
\unitlength 0.90mm
\linethickness{0.4pt}
\begin{picture}(140.00,110.00)
\put(140.00,100.00){\vector(1,0){0.2}}
\emline{10.00}{100.00}{1}{140.00}{100.00}{2}
\emline{120.00}{125.00}{3}{120.92}{122.58}{4}
\emline{120.92}{122.58}{5}{121.75}{120.17}{6}
\emline{121.75}{120.17}{7}{122.48}{117.75}{8}
\emline{122.48}{117.75}{9}{123.12}{115.33}{10}
\emline{123.12}{115.33}{11}{123.67}{112.92}{12}
\emline{123.67}{112.92}{13}{124.12}{110.50}{14}
\emline{124.12}{110.50}{15}{124.48}{108.08}{16}
\emline{124.48}{108.08}{17}{124.74}{105.67}{18}
\emline{124.74}{105.67}{19}{124.92}{103.25}{20}
\emline{124.92}{103.25}{21}{124.99}{100.83}{22}
\emline{124.99}{100.83}{23}{124.98}{98.42}{24}
\emline{124.98}{98.42}{25}{124.87}{96.00}{26}
\emline{124.87}{96.00}{27}{124.67}{93.58}{28}
\emline{124.67}{93.58}{29}{124.38}{91.17}{30}
\emline{124.38}{91.17}{31}{123.99}{88.75}{32}
\emline{123.99}{88.75}{33}{123.51}{86.33}{34}
\emline{123.51}{86.33}{35}{122.93}{83.92}{36}
\emline{122.93}{83.92}{37}{122.26}{81.50}{38}
\emline{122.26}{81.50}{39}{121.50}{79.08}{40}
\emline{121.50}{79.08}{41}{120.00}{75.00}{42}
\emline{30.00}{75.00}{43}{29.08}{77.42}{44}
\emline{29.08}{77.42}{45}{28.25}{79.83}{46}
\emline{28.25}{79.83}{47}{27.52}{82.25}{48}
\emline{27.52}{82.25}{49}{26.88}{84.67}{50}
\emline{26.88}{84.67}{51}{26.33}{87.08}{52}
\emline{26.33}{87.08}{53}{25.88}{89.50}{54}
\emline{25.88}{89.50}{55}{25.52}{91.92}{56}
\emline{25.52}{91.92}{57}{25.26}{94.33}{58}
\emline{25.26}{94.33}{59}{25.08}{96.75}{60}
\emline{25.08}{96.75}{61}{25.01}{99.17}{62}
\emline{25.01}{99.17}{63}{25.02}{101.58}{64}
\emline{25.02}{101.58}{65}{25.13}{104.00}{66}
\emline{25.13}{104.00}{67}{25.33}{106.42}{68}
\emline{25.33}{106.42}{69}{25.62}{108.83}{70}
\emline{25.62}{108.83}{71}{26.01}{111.25}{72}
\emline{26.01}{111.25}{73}{26.49}{113.67}{74}
\emline{26.49}{113.67}{75}{27.07}{116.08}{76}
\emline{27.07}{116.08}{77}{27.74}{118.50}{78}
\emline{27.74}{118.50}{79}{28.50}{120.92}{80}
\emline{28.50}{120.92}{81}{30.00}{125.00}{82}
\put(17.00,57.33){\makebox(0,0)[lc]
{\small Fig.2: A scheme of conversion of an electromagnetic laser beam exited}}
\put(20.67,52.33){\makebox(0,0)[lc]
{\small in an open resonator into gravitational one; $M_1$, $M_2$: Mirrors;}}
\put(20.67,47.33){\makebox(0,0)[lc]
{\small $N$, $S$: magnetic poles; $BE$: Beam Envelops of an electromagnetic and}}
\put(20.67,42.33){\makebox(0,0)[lc]
{\small gravitational radiations.}}
\emline{25.67}{110.00}{83}{29.15}{109.32}{84}
\emline{29.15}{109.32}{85}{32.63}{108.69}{86}
\emline{32.63}{108.69}{87}{36.11}{108.11}{88}
\emline{36.11}{108.11}{89}{39.59}{107.58}{90}
\emline{39.59}{107.58}{91}{43.07}{107.10}{92}
\emline{43.07}{107.10}{93}{46.55}{106.66}{94}
\emline{46.55}{106.66}{95}{50.03}{106.28}{96}
\emline{50.03}{106.28}{97}{53.51}{105.95}{98}
\emline{53.51}{105.95}{99}{56.98}{105.67}{100}
\emline{56.98}{105.67}{101}{60.46}{105.43}{102}
\emline{60.46}{105.43}{103}{63.94}{105.25}{104}
\emline{63.94}{105.25}{105}{67.42}{105.12}{106}
\emline{67.42}{105.12}{107}{70.90}{105.03}{108}
\emline{70.90}{105.03}{109}{74.38}{105.00}{110}
\emline{74.38}{105.00}{111}{77.86}{105.02}{112}
\emline{77.86}{105.02}{113}{81.34}{105.08}{114}
\emline{81.34}{105.08}{115}{84.82}{105.20}{116}
\emline{84.82}{105.20}{117}{88.30}{105.36}{118}
\emline{88.30}{105.36}{119}{91.78}{105.58}{120}
\emline{91.78}{105.58}{121}{95.26}{105.84}{122}
\emline{95.26}{105.84}{123}{98.74}{106.16}{124}
\emline{98.74}{106.16}{125}{102.22}{106.52}{126}
\emline{102.22}{106.52}{127}{105.70}{106.94}{128}
\emline{105.70}{106.94}{129}{109.18}{107.40}{130}
\emline{109.18}{107.40}{131}{112.66}{107.91}{132}
\emline{112.66}{107.91}{133}{116.14}{108.48}{134}
\emline{116.14}{108.48}{135}{119.61}{109.09}{136}
\emline{119.61}{109.09}{137}{124.33}{110.00}{138}
\emline{124.33}{90.00}{139}{120.85}{90.68}{140}
\emline{120.85}{90.68}{141}{117.37}{91.31}{142}
\emline{117.37}{91.31}{143}{113.89}{91.89}{144}
\emline{113.89}{91.89}{145}{110.41}{92.42}{146}
\emline{110.41}{92.42}{147}{106.93}{92.90}{148}
\emline{106.93}{92.90}{149}{103.45}{93.34}{150}
\emline{103.45}{93.34}{151}{99.97}{93.72}{152}
\emline{99.97}{93.72}{153}{96.49}{94.05}{154}
\emline{96.49}{94.05}{155}{93.02}{94.33}{156}
\emline{93.02}{94.33}{157}{89.54}{94.57}{158}
\emline{89.54}{94.57}{159}{86.06}{94.75}{160}
\emline{86.06}{94.75}{161}{82.58}{94.88}{162}
\emline{82.58}{94.88}{163}{79.10}{94.97}{164}
\emline{79.10}{94.97}{165}{75.62}{95.00}{166}
\emline{75.62}{95.00}{167}{72.14}{94.98}{168}
\emline{72.14}{94.98}{169}{68.66}{94.92}{170}
\emline{68.66}{94.92}{171}{65.18}{94.80}{172}
\emline{65.18}{94.80}{173}{61.70}{94.64}{174}
\emline{61.70}{94.64}{175}{58.22}{94.42}{176}
\emline{58.22}{94.42}{177}{54.74}{94.16}{178}
\emline{54.74}{94.16}{179}{51.26}{93.84}{180}
\emline{51.26}{93.84}{181}{47.78}{93.48}{182}
\emline{47.78}{93.48}{183}{44.30}{93.06}{184}
\emline{44.30}{93.06}{185}{40.82}{92.60}{186}
\emline{40.82}{92.60}{187}{37.34}{92.09}{188}
\emline{37.34}{92.09}{189}{33.86}{91.52}{190}
\emline{33.86}{91.52}{191}{30.39}{90.91}{192}
\emline{30.39}{90.91}{193}{25.67}{90.00}{194}
\emline{50.00}{135.00}{195}{50.00}{125.67}{196}
\emline{50.00}{125.67}{197}{100.00}{125.67}{198}
\emline{100.00}{125.67}{199}{100.00}{134.67}{200}
\emline{100.00}{134.67}{201}{100.00}{134.67}{202}
\emline{100.00}{65.00}{203}{100.00}{75.00}{204}
\emline{100.00}{75.00}{205}{50.00}{75.00}{206}
\emline{50.00}{75.00}{207}{50.00}{65.00}{208}
\emline{50.00}{65.00}{209}{50.00}{65.00}{210}
\put(120.00,130.67){\makebox(0,0)[cc]{$M2$}}
\put(30.00,130.67){\makebox(0,0)[cc]{$M1$}}
\put(75.00,69.67){\makebox(0,0)[cc]{$S$}}
\put(75.00,130.00){\makebox(0,0)[cc]{$N$}}
\put(106.33,110.00){\makebox(0,0)[cc]{$BE$}}
\put(138.33,115.67){\vector(4,3){0.2}}
\emline{124.33}{110.00}{211}{126.69}{110.40}{212}
\emline{126.69}{110.40}{213}{128.99}{110.98}{214}
\emline{128.99}{110.98}{215}{131.23}{111.74}{216}
\emline{131.23}{111.74}{217}{133.39}{112.68}{218}
\emline{133.39}{112.68}{219}{135.50}{113.80}{220}
\emline{135.50}{113.80}{221}{138.33}{115.67}{222}
\put(138.33,84.33){\vector(4,-3){0.2}}
\emline{124.33}{90.00}{223}{127.57}{89.16}{224}
\emline{127.57}{89.16}{225}{130.42}{88.30}{226}
\emline{130.42}{88.30}{227}{132.88}{87.42}{228}
\emline{132.88}{87.42}{229}{134.96}{86.53}{230}
\emline{134.96}{86.53}{231}{136.63}{85.63}{232}
\emline{136.63}{85.63}{233}{138.33}{84.33}{234}
\put(138.33,97.00){\makebox(0,0)[cc]{$y$}}
\end{picture}

             \end{center}
             \vspace{-37mm}

       \section{Detectors of the gravitational radiation}

The schemes of detectors of artificial sources of gravitational
waves can be based both on the coherent
emission of the electromagnetic radiation by charged particle beams or
another charged formations accelerated by the gravitational
waves and on an inverse conversion of gravitons into photons in the
transverse magnetic field. Multilayer mirrors can
be created from charged electron and ion beams in storage rings (moving
multilayer mirrors) \cite{microb} or from transparent charged
dielectric layers placed along the way of the gravitational beam.

\subsection{\it Detectors based on coherent conversion of gravitons into
photons by ion bunches (backward Gravi-Compton scattering)}

The conversion cross section of the graviton to photon on a charged ion
is determined by the equation

                \begin{equation}       
           \sigma = \sigma _i {C_{\gamma}GM_i^2\over e^2} =
                \sigma _e {C_{\gamma}Gm_e^2(n^{+})^2\over e^2}
                \end{equation}
where $\sigma_i = \sigma _e(m_e/M_i)^2(n^{+})^2$ is the ion scattering
cross section, $\sigma _e = {8\pi r_e^2/3} \simeq 6.65 \cdot
10^{-25}$\,$cm^2$ Thomson scattering cross section. Notice that the
cross section (10) does not depend on mass of particle but depend on
ion charge state $n^{+}$.

In the case of prebunched ion beam in the storage ring (moving
multilayer ion mirror) the coherent cross section $\sigma_{coh}$ per
one ion will be the cross section (10) multiplied by the coherence
factor which is equal $6M_bN_1$ for the point like bunches (transverse
dimensions less then $\lambda \gamma $, longitudinal dimensions less
then $\lambda $), where $M_b$ is the number of bunches in the straight
section determined by the length of the straight section $L$ and the
wavelength $\lambda _{rf}$ of the radio frequency system of the storage
ring ($M_b = L/\lambda _{rf}$), $N_1 = i\lambda _{rf}/ecn^{+} \simeq
2.08 \cdot 10^{8} \lambda _{rf}i[A]/n^{+}$ the number of ions in one
ion bunch, $\lambda = \lambda _{rf}/4\gamma ^2$ is the wavelength of
the backward scattered electromagnetic radiation (process: graviton +
ion $\to$ photon + ion), $\gamma $ the relativistic factor of ions. The
coefficient 6 includes the coefficient 2 which appeared from the fact
that in the relativistic case each graviton will meat 2$M_b$ ion
bunches for the time of crossing of the straight section and
coefficient 3 which take into account the coherence between bunches. It
follows that in this case the coherent cross section is equal

                \begin{equation}       
           \sigma _{coh} = 6M_bN_1\sigma \simeq 6.47\cdot
           10^{-58}M_bi[A]\lambda _{rf}[cm]/n^{+}.  \end{equation}

The luminosity of the storage ring under consideration

                \begin{equation}       
           L_{gi} = {2iL\dot N^{gr}\over ecS_{gr}}
           \simeq 4.17\cdot 10^{8}i[A]L[sm]/S_{gr}[cm^2]. \end{equation}
where $S_{gr}$ is the area of the gravitational beam.

In the case of i = 1kA, $\lambda _{rf}$ = 1 cm, ($N_1 = 2.08 \cdot
10^{11}$), $M_b = 10^4$ (L = 100m), $n^{+} =1$, $S_{gr} = 10^4 cm^2$:
$\sigma _ {coh} \simeq 6.47\cdot 10^{-51}$ cm$^2$, $L_{gi} = 4.17\cdot
10^{11}$ cm$^{-2}$ s$^{-1}$, and the flow of the backward scattered
photons $\dot N^{bs} = 2.7 \cdot 10^{-39}$ ph/s. The energy of
scattered radiation will be determined by the energy of the dedicated
storage ring and at ($\gamma \simeq 2\div 5$) will lay in the
submillimeter and more hard regions. This scheme permit to increase the
energy of the emitted photons the more the higher the energy of the
storage ring. High quality ($Q \gg 1$) open resonator can be used to
increase the low of photons $Q^2$ times inside and $Q$ times outside of
the resonator (extracted).

   \subsection{\it Detectors based on conversion of gravitons into
photons by a charged multilayer mirror}

The coefficient of reflection of electromagnetic waves by a thin layer
of electrons is equal $k_r = r_e^2N_{\sigma}^2\lambda ^2$, where
$N_{\sigma }$ is the electric charge surface density, $\lambda $ the
wavelength of the falling radiation \cite{2landau}. When electrons are
arranged in $M_l$ layers then in this case the maximal coefficient of
reflection will be $M_l^2$ times higher.

By analogy the gravitational waves will lead to an acceleration of
electrons of the multilayer mirror and hence to emission of
electromagnetic waves. The converted electromagnetic waves will
propagate the same direction as the reflected electromagnetic waves.
The conversion coefficient of the gravitational radiation to
electromagnetic one in this case will be less then the reflection
coefficient of the electromagnetic waves in accordance with the
coefficient $C_{\gamma} Gm_e^2/e^2$:

                \begin{equation}       
     k_{ml} = C_{\gamma }Gm_e^2r_e^2 N_{\sigma}^2\lambda ^2 M_l^2/e^2.
                 \end{equation}

On practice transparent dielectric media can be cut off on thin flat
layers. These layers can be charged from both sides by electric charges
of the opposite sign. They can be arranged such a way that the surfaces
with identical signs of charges was brought together as shown on Fig.3.

             \vspace{5mm}
             \begin{center}
\special{em:linewidth 0.4pt}
\unitlength 0.90mm
\linethickness{0.4pt}
\begin{picture}(140.00,131.67)
\put(140.00,100.00){\vector(1,0){0.2}}
\emline{10.00}{100.00}{1}{140.00}{100.00}{2}
\emline{120.00}{125.00}{3}{120.92}{122.58}{4}
\emline{120.92}{122.58}{5}{121.75}{120.17}{6}
\emline{121.75}{120.17}{7}{122.48}{117.75}{8}
\emline{122.48}{117.75}{9}{123.12}{115.33}{10}
\emline{123.12}{115.33}{11}{123.67}{112.92}{12}
\emline{123.67}{112.92}{13}{124.12}{110.50}{14}
\emline{124.12}{110.50}{15}{124.48}{108.08}{16}
\emline{124.48}{108.08}{17}{124.74}{105.67}{18}
\emline{124.74}{105.67}{19}{124.92}{103.25}{20}
\emline{124.92}{103.25}{21}{124.99}{100.83}{22}
\emline{124.99}{100.83}{23}{124.98}{98.42}{24}
\emline{124.98}{98.42}{25}{124.87}{96.00}{26}
\emline{124.87}{96.00}{27}{124.67}{93.58}{28}
\emline{124.67}{93.58}{29}{124.38}{91.17}{30}
\emline{124.38}{91.17}{31}{123.99}{88.75}{32}
\emline{123.99}{88.75}{33}{123.51}{86.33}{34}
\emline{123.51}{86.33}{35}{122.93}{83.92}{36}
\emline{122.93}{83.92}{37}{122.26}{81.50}{38}
\emline{122.26}{81.50}{39}{121.50}{79.08}{40}
\emline{121.50}{79.08}{41}{120.00}{75.00}{42}
\emline{30.00}{75.00}{43}{29.08}{77.42}{44}
\emline{29.08}{77.42}{45}{28.25}{79.83}{46}
\emline{28.25}{79.83}{47}{27.52}{82.25}{48}
\emline{27.52}{82.25}{49}{26.88}{84.67}{50}
\emline{26.88}{84.67}{51}{26.33}{87.08}{52}
\emline{26.33}{87.08}{53}{25.88}{89.50}{54}
\emline{25.88}{89.50}{55}{25.52}{91.92}{56}
\emline{25.52}{91.92}{57}{25.26}{94.33}{58}
\emline{25.26}{94.33}{59}{25.08}{96.75}{60}
\emline{25.08}{96.75}{61}{25.01}{99.17}{62}
\emline{25.01}{99.17}{63}{25.02}{101.58}{64}
\emline{25.02}{101.58}{65}{25.13}{104.00}{66}
\emline{25.13}{104.00}{67}{25.33}{106.42}{68}
\emline{25.33}{106.42}{69}{25.62}{108.83}{70}
\emline{25.62}{108.83}{71}{26.01}{111.25}{72}
\emline{26.01}{111.25}{73}{26.49}{113.67}{74}
\emline{26.49}{113.67}{75}{27.07}{116.08}{76}
\emline{27.07}{116.08}{77}{27.74}{118.50}{78}
\emline{27.74}{118.50}{79}{28.50}{120.92}{80}
\emline{28.50}{120.92}{81}{30.00}{125.00}{82}
\put(17.00,57.33){\makebox(0,0)[lc]
{\small Fig.3: A scheme of conversion of a gravitational beam into electro-}}
\put(25.67,52.33){\makebox(0,0)[lc]
{\small magnetic one; $M_1$, $M_2$: Mirrors; $BE$: Beam Envelop of an }}
\put(25.67,47.33){\makebox(0,0)[lc]
{\small electromagnetic radiation; M3: Mirror for extraction of radiation. }}
\emline{25.67}{110.00}{83}{29.15}{109.32}{84}
\emline{29.15}{109.32}{85}{32.63}{108.69}{86}
\emline{32.63}{108.69}{87}{36.11}{108.11}{88}
\emline{36.11}{108.11}{89}{39.59}{107.58}{90}
\emline{39.59}{107.58}{91}{43.07}{107.10}{92}
\emline{43.07}{107.10}{93}{46.55}{106.66}{94}
\emline{46.55}{106.66}{95}{50.03}{106.28}{96}
\emline{50.03}{106.28}{97}{53.51}{105.95}{98}
\emline{53.51}{105.95}{99}{56.98}{105.67}{100}
\emline{56.98}{105.67}{101}{60.46}{105.43}{102}
\emline{60.46}{105.43}{103}{63.94}{105.25}{104}
\emline{63.94}{105.25}{105}{67.42}{105.12}{106}
\emline{67.42}{105.12}{107}{70.90}{105.03}{108}
\emline{70.90}{105.03}{109}{74.38}{105.00}{110}
\emline{74.38}{105.00}{111}{77.86}{105.02}{112}
\emline{77.86}{105.02}{113}{81.34}{105.08}{114}
\emline{81.34}{105.08}{115}{84.82}{105.20}{116}
\emline{84.82}{105.20}{117}{88.30}{105.36}{118}
\emline{88.30}{105.36}{119}{91.78}{105.58}{120}
\emline{91.78}{105.58}{121}{95.26}{105.84}{122}
\emline{95.26}{105.84}{123}{98.74}{106.16}{124}
\emline{98.74}{106.16}{125}{102.22}{106.52}{126}
\emline{102.22}{106.52}{127}{105.70}{106.94}{128}
\emline{105.70}{106.94}{129}{109.18}{107.40}{130}
\emline{109.18}{107.40}{131}{112.66}{107.91}{132}
\emline{112.66}{107.91}{133}{116.14}{108.48}{134}
\emline{116.14}{108.48}{135}{119.61}{109.09}{136}
\emline{119.61}{109.09}{137}{124.33}{110.00}{138}
\emline{124.33}{90.00}{139}{120.85}{90.68}{140}
\emline{120.85}{90.68}{141}{117.37}{91.31}{142}
\emline{117.37}{91.31}{143}{113.89}{91.89}{144}
\emline{113.89}{91.89}{145}{110.41}{92.42}{146}
\emline{110.41}{92.42}{147}{106.93}{92.90}{148}
\emline{106.93}{92.90}{149}{103.45}{93.34}{150}
\emline{103.45}{93.34}{151}{99.97}{93.72}{152}
\emline{99.97}{93.72}{153}{96.49}{94.05}{154}
\emline{96.49}{94.05}{155}{93.02}{94.33}{156}
\emline{93.02}{94.33}{157}{89.54}{94.57}{158}
\emline{89.54}{94.57}{159}{86.06}{94.75}{160}
\emline{86.06}{94.75}{161}{82.58}{94.88}{162}
\emline{82.58}{94.88}{163}{79.10}{94.97}{164}
\emline{79.10}{94.97}{165}{75.62}{95.00}{166}
\emline{75.62}{95.00}{167}{72.14}{94.98}{168}
\emline{72.14}{94.98}{169}{68.66}{94.92}{170}
\emline{68.66}{94.92}{171}{65.18}{94.80}{172}
\emline{65.18}{94.80}{173}{61.70}{94.64}{174}
\emline{61.70}{94.64}{175}{58.22}{94.42}{176}
\emline{58.22}{94.42}{177}{54.74}{94.16}{178}
\emline{54.74}{94.16}{179}{51.26}{93.84}{180}
\emline{51.26}{93.84}{181}{47.78}{93.48}{182}
\emline{47.78}{93.48}{183}{44.30}{93.06}{184}
\emline{44.30}{93.06}{185}{40.82}{92.60}{186}
\emline{40.82}{92.60}{187}{37.34}{92.09}{188}
\emline{37.34}{92.09}{189}{33.86}{91.52}{190}
\emline{33.86}{91.52}{191}{30.39}{90.91}{192}
\emline{30.39}{90.91}{193}{25.67}{90.00}{194}
\put(138.33,97.00){\makebox(0,0)[cc]{$y$}}
\emline{55.00}{80.00}{195}{55.00}{120.00}{196}
\emline{63.00}{80.00}{197}{63.00}{120.00}{198}
\emline{71.00}{80.00}{199}{71.00}{120.00}{200}
\emline{79.00}{80.00}{201}{79.00}{120.00}{202}
\emline{87.00}{80.00}{203}{87.00}{120.00}{204}
\emline{95.00}{80.00}{205}{95.00}{120.00}{206}
\emline{55.00}{120.00}{207}{95.00}{120.00}{208}
\emline{95.00}{80.00}{209}{55.00}{80.00}{210}
\put(75.33,124.67){\makebox(0,0)[cc]{TM}}
\put(120.00,130.67){\makebox(0,0)[cc]{M2}}
\put(30.00,130.67){\makebox(0,0)[cc]{M1}}
\put(111.67,114.33){\makebox(0,0)[cc]{ME}}
\put(56.67,117.00){\makebox(0,0)[cc]{+}}
\put(72.67,117.00){\makebox(0,0)[cc]{+}}
\put(88.67,117.00){\makebox(0,0)[cc]{+}}
\put(56.67,109.67){\makebox(0,0)[cc]{{\small +}}}
\put(56.67,102.33){\makebox(0,0)[cc]{+}}
\put(56.67,95.00){\makebox(0,0)[cc]{+}}
\put(56.67,89.00){\makebox(0,0)[cc]{+}}
\put(56.67,83.00){\makebox(0,0)[cc]{+}}
\put(72.67,109.60){\makebox(0,0)[cc]{+}}
\put(72.67,102.00){\makebox(0,0)[cc]{+}}
\put(72.67,95.00){\makebox(0,0)[cc]{+}}
\put(72.67,89.00){\makebox(0,0)[cc]{+}}
\put(72.67,83.00){\makebox(0,0)[cc]{+}}
\put(88.67,109.60){\makebox(0,0)[cc]{+}}
\put(88.67,102.00){\makebox(0,0)[cc]{+}}
\put(88.67,95.00){\makebox(0,0)[cc]{+}}
\put(88.67,89.00){\makebox(0,0)[cc]{+}}
\put(88.67,83.33){\makebox(0,0)[cc]{+}}
\put(61.33,117.33){\makebox(0,0)[cc]{--}}
\put(61.33,110.00){\makebox(0,0)[cc]{--}}
\put(61.33,102.67){\makebox(0,0)[cc]{--}}
\put(61.33,95.33){\makebox(0,0)[cc]{--}}
\put(61.33,88.00){\makebox(0,0)[cc]{--}}
\put(61.33,80.67){\makebox(0,0)[cc]{--}}
\put(64.67,117.33){\makebox(0,0)[cc]{--}}
\put(64.67,110.33){\makebox(0,0)[cc]{--}}
\put(64.67,103.33){\makebox(0,0)[cc]{--}}
\put(64.67,96.33){\makebox(0,0)[cc]{--}}
\put(64.67,89.33){\makebox(0,0)[cc]{--}}
\put(64.67,82.33){\makebox(0,0)[cc]{--}}
\put(77.33,117.67){\makebox(0,0)[cc]{--}}
\put(77.33,110.67){\makebox(0,0)[cc]{--}}
\put(77.33,103.67){\makebox(0,0)[cc]{--}}
\put(77.33,96.67){\makebox(0,0)[cc]{--}}
\put(77.33,89.67){\makebox(0,0)[cc]{--}}
\put(77.33,82.67){\makebox(0,0)[cc]{--}}
\put(69.33,117.33){\makebox(0,0)[cc]{+}}
\put(69.33,110.33){\makebox(0,0)[cc]{+}}
\put(69.33,103.33){\makebox(0,0)[cc]{+}}
\put(69.33,96.33){\makebox(0,0)[cc]{+}}
\put(69.33,89.33){\makebox(0,0)[cc]{+}}
\put(69.33,82.33){\makebox(0,0)[cc]{+}}
\put(80.67,118.00){\makebox(0,0)[cc]{--}}
\put(80.67,111.00){\makebox(0,0)[cc]{--}}
\put(80.67,104.00){\makebox(0,0)[cc]{--}}
\put(80.67,97.00){\makebox(0,0)[cc]{--}}
\put(80.67,90.00){\makebox(0,0)[cc]{--}}
\put(80.67,83.00){\makebox(0,0)[cc]{--}}
\put(85.33,117.67){\makebox(0,0)[cc]{+}}
\put(85.33,110.67){\makebox(0,0)[cc]{+}}
\put(85.33,103.67){\makebox(0,0)[cc]{+}}
\put(85.33,96.67){\makebox(0,0)[cc]{+}}
\put(85.33,89.67){\makebox(0,0)[cc]{+}}
\put(85.33,82.67){\makebox(0,0)[cc]{+}}
\put(93.33,117.33){\makebox(0,0)[cc]{--}}
\put(93.33,110.00){\makebox(0,0)[cc]{--}}
\put(93.33,102.67){\makebox(0,0)[cc]{--}}
\put(93.33,95.33){\makebox(0,0)[cc]{--}}
\put(93.33,88.00){\makebox(0,0)[cc]{--}}
\put(93.33,80.67){\makebox(0,0)[cc]{--}}
\emline{113.00}{93.00}{211}{114.67}{90.00}{212}
\put(115.33,86.00){\makebox(0,0)[cc]{M3}}
\put(96.67,66.67){\vector(-2,-3){0.2}}
\emline{113.67}{91.33}{213}{96.67}{66.67}{214}
\end{picture}

             \end{center}
             \vspace{-40mm}

The propagating in the direction of the axis "y" gravitational
radiation will be converted into electromagnetic radiation by charged
layers. The open resonator with a quality $Q$ will stimulate the
conversion process $Q$ times. As a result the flow of photons will be
increased $Q^2$ times inside and $Q$ times outside of the system.
The phases of the waves emitted by layers with opposite polarities will
differ on $\pi$. The velocity of gravitational waves are equal to
light velocity in any case but the velocity of the electromagnetic
waves will depend on the refraction index of the media. That is why we
can choose the distance between layers such a way that the emitted
electromagnetic waves will be in phase forward or backward
directions.

In this scheme the surface density of electric charge will be limited
by the dielectric breakdown electric field strength $E_{br}$ ($E_{br}
\leq 6\cdot 10^3$ or $\leq 2\cdot 10^6$ [V/cm])

                \begin{equation}       
N_{\sigma} = {E_{br}\over 2\pi e} \simeq 1.1\cdot 10^6 E_{br}[V/cm].
                 \end{equation}

According to (13), (14) the efficiency of the system

                \begin{equation}       
         {\dot N_r^{em}\over \dot N^{gr}} = k_{ml}\cdot Q^2 \simeq 7.51
                 \cdot 10^{-56}\lambda_{rf}^2M_l^2E_{br}^2Q^2.
                \end{equation}
where $\dot N_r^{em}$ is the flow of the converted photons stored
inside the resonator. The flow of photons extracted from the resonator
is $Q$ times less.

In the case of $\lambda _{rf} = 1 cm$, $M_l = 10^5$ ($L \simeq $
1\,km), $Q = 10^9$, $E_{br}$ = 2MV/cm, the efficiency ${\dot
N^{em}_r/\dot N^{gr}} = 3 \cdot 10^{-15}$. The flow of photons $\dot
N^{em}_r = 3 \cdot 10^{-12}$ ph/s when $\dot N^{gr}$ = $10^3$ gr/s. This
value is much higher then in the previous one.

  \subsection{\it Detectors based on conversion of gravitons into
photons by  charged mirrors of an open resonator}

If in the scheme of Fig.3 the media in a resonator is absent but
mirrors $M1$ and $M2$ are charged then such system will be excited. The
efficiency of such system will be much less then in the previous case.
Detectors based on the emission of charged capacitor with transparent
for radiation plates in vacuum was considered in \cite{lupanov}.

       \subsection{\it Detectors based on conversion of gravitons into
         photons in a transverse magnetic field}

The conversion of the gravitational radiation into electromagnetic one
can be produced in the conversion scheme which is reverse one to the
scheme presented in the Fig.2 \cite{mitskevich}, \cite{boccal}. The
gravitational radiation in this scheme propagates in the direction of
axis "y" and is converted to an electromagnetic radiation in the
transverse magnetic field and in the electromagnetic field stored in
the open resonator. The conversion coefficient in this case is

                \begin{equation}       
          p_{g\to \gamma } = {\dot N_r^{em}\over \dot N^{gr}} =
                {G\over c^4}Q^2L^2B_{\perp}^2.
                 \end{equation}
where $\dot N_r^{em}$ is the flow of photons stored in the resonator.

The quality of a resonator $Q^2$ in (16) takes into account stimulation
of the conversion of the gravitational radiation by the resonator. At
that the extraction of the energy from the gravitational wave will
be increased $Q$ times. The number of the photons stored in the
resonator for the time $\Delta t > \tau$ is

                 \begin{equation}       
     \Delta N^{em} = \dot N_r^{em}T = {2G\over c^5}Q^2L^2L_mB
       _{\perp} ^2\dot N^{gr},  \end{equation}
where $\tau = Q\,T$ is the rising time (increment) of this system, $T
= 2L_m/c$ double time of the light passage between mirrors of the
resonator.

In the case of $B_{\perp} = 10^5$\,G, $L_m \simeq L = 1$\,km, $P^{em}_a
= 10^2$\,W, $Q = 10^9$, $P^{em}_r = 10^{11}$\,W, $P^{gr} = 8.26 \cdot
10^{-19}$\,W, $\dot N^{gr} = 4\cdot 10^4$\,gr/s, $\hbar \omega _{gr} =
1.25 \cdot 10^{-4}$\,eV or $\lambda $ = 1cm according to (16) the value
${\dot N^{em}_r/ \dot N^{gr}} = 8.26\cdot 10^{-12}$, the flow of
converted photons in the resonator $\dot N^{em}_r \simeq 3.3\cdot
10^{-7}$ ph/s, the number of the photons stored in the resonator for
the time $\Delta t > \tau = 6.7\cdot 10^3$\,ms is $\Delta N^{em} =
2.2\cdot 10^{-8}$.  This is very small value. Of cause we can take the
length of the resonator, the value of the magnetic field and the flow
of the gravitons much higher \cite{chiarello}, \cite{chen}. In this
case the flow of gravitons can achieve an accessible value (some
photons per a day).

          \subsection{\it Stimulation of the conversion of gravitons
                   into photons by an open resonator}

When an oscillator emit in an open resonator an electromagnetic
wavepacket then a part of this wavepacket will be stored in transverse
and longitudinal modes. After reflections from the mirrors of the
resonator the amplitude $E_{\alpha 0}$ of the emitted wavepacket will
be decreased $(r_lr_r)^n$ times, where $r_i^2$ are the reflectivity
coefficients of the left $l$ and right $r$ mirrors accordingly, $n$ the
number of reflections from the pair of mirrors. When oscillator emit a
train of wavepackets with a period $T$ then after $N$ reflections of 
the first wavepacket the amplitude of the total electric field strength 
of wavepackets stored in the resonator will be determined by the 
expression

                 \begin{equation}       
      E_{\alpha}^{\Sigma} = \sum _{n=0}^N E_{\alpha 0}(r_lr_r)^n =
      E_{\alpha 0}Q(1 - (r_lr_r)^N)
         \end{equation}
where the value $Q = 1/(1-r_l r_r)$ is the quality of the resonator.
The value $N \simeq t/T$ is the number of reflections of the first 
wavepacket for the time interval $t$, the value $(r_lr_r)^N \simeq 
\exp[-t/2\tau]$, where $\tau = - T/\ln{(r_lr_r)^2}| _{Q\gg 1} \simeq 
T\,Q/2$.

The energy stored in the resonator will decay by the low: $\varepsilon
^{em} = \varepsilon ^{em}_0\exp(-t/\tau)$.

In the case of the main transverse mode in the resonator $TEM_{00}$ the
density of the electromagnetic energy in the mode will have Gaussian
form. The power of the electromagnetic radiation in the mode will be
determined by the product of the Poynting's vector for the intensity $I
= (c/8\pi)E_m^2$ and the area of the mode $\pi \sigma _0^2$, where
$E_m$ is the amplitude of the linear polarized wave. In practical
unites $I = (1/240\pi) E_m^2 \simeq 1.33\cdot 10^{-3} E_m^2$\,[W/V$^2
$]. The maximum average intensities of the radiation reflected from the
mirrors $I = 1.33\cdot 10^7 \,W/cm^2$ and $I = 1.33\cdot 10^{9}$ \,W/cm
$^2$ are limited by heating of mirrors and correspond to the electric
field strengths accordingly $E_m \simeq 10^5[$V/cm] for the case of
warm metal mirrors and to $E_m \simeq 10^6[$V/cm] for the
superconducting mirrors in a \,cm\, wavelength region.

     \section{On a possibility of laboratory gravitational experiments}

The combined symmetrical scheme of an experiment based on
double conversion of electromagnetic to gravitational radiation and
back is presented on Fig.4. In this scheme two equal semi-confocal
resonators of the length $L_m$ each are presented by mirrors $M1$ -
$M3$. $M3$ is a thick flat mirror, which is simultaneously the absorber
of the powerful electromagnetic radiation exited in the left resonator.

In this case the ratio of the reactive power of the electromagnetic
radiation stored into the right resonator to the active power of
electromagnetic radiation supplying the left resonator and the number
of photons stored in the right resonator are

                \begin{equation}       
      p_{\gamma \to g \to \gamma } = {P^{em}_{r\,r}\over
                P^{em}_{a\,l}} = {\dot N^{em}
        _{r\,r}\over \dot N^{em}_{a\,l}} = {G^2\over
            c^8}Q_lQ_r^2L^4 B_{\perp}^4,   \end{equation}

                \begin{equation}       
     \Delta N^{em}_r = p_{\gamma \to g \to \gamma }\dot N^{em}_{a\,l}\,T,
                \end{equation}
where $G^2/c^8 \simeq 6.78 \cdot 10^{-99}$, $Q_l$ and $Q_r$ are the
qualities of the left and right resonators accordingly.

   \vskip 5mm       {\it \bf Example 3.}\hskip 4mm
Let us the qualities of the superconducting resonators are $Q_1 = Q_2 =
10^9$, the value of the transverse magnetic fields $B_{\perp} =
10^6$ Gs, the lengths of the magnet $L = 2\cdot 10^6$ cm, the distance
between mirrors $M1$ - $M3$ and $M3$ $M2$  $L_m \simeq L/2 = 10^6$ cm,
$\hbar \omega _{gr} = 1.25 \cdot 10^{-4}$\,eV ($\lambda $ = 1cm), the
intensity of the electromagnetic wave stored in the resonator $I =
1.33\cdot 10^9$[W/cm$^2$], the active power supplying the left
resonator $P^{em}_{a\,l} = 6.62\cdot 10^{5}$W.

In this case the reactive power stored in the resonator $P^{em}_{a\,l}
= 6.62 \cdot 10^{14}$W, the flow of photons from the laser supplying
the left resonator $\dot N^{em}_{a\,l} =3.31 \cdot 10^{28}$\,ph/s, the
flow of photons stored in the left resonator $N^{em}_{a\,l} =3.31 \cdot
10^{37}$\,ph/s, and according to (19) the ratio $p_{\gamma \to g \to
\gamma} = 6.78\cdot 10^{-24}$, $\sigma _0 = 398$\,cm,  $T = (2/3)\cdot
10^{-5}$\,s, $\tau = (2/3)\cdot 10^4 $\,s$\simeq 2$ hours, $\Delta
N^{em} = 0.748\,gr$, the flow of photons $\dot N^{em}_{r,r} = 2.24\cdot
10^{-4} $\,ph/s or $\sim$ 1\,ph/hour. At that in the right resonator
the flow of gravitons is $\dot N^{gr}_r = 2.72\cdot 10^{12}$\,gr/s.

The number of photons stored in the right resonator $\Delta N^{em}
\simeq 1$ or the flow of photons $\dot N^{em}_{r,r} \sim $ 1\,ph/hour
are not so small to reject the possibility of the gravitational
experiments in principle. The value of the magnetic field $\sim
10^6$\,G is not reached now on experiment but it does not exceed the
theoretical limit. The quality of the open resonator $Q\sim 10^{11}$
was achieved in the open resonator in the decimeter wavelength
region. The diameter of the superconducting mirror $\sim$\,10m, the
volume of the superconducting magnet ($10^6$\,m$^3$), and the average
power of a laser or a free-particle laser $\sim 1$\,MW are very huge
and expensive but capable of execution.

\vspace{3mm} \begin{center}
\special{em:linewidth 0.4pt}
\unitlength 0.90mm
\linethickness{0.4pt}
\begin{picture}(140.00,135.00)
\put(140.00,100.00){\vector(1,0){0.2}}
\emline{10.00}{100.00}{1}{140.00}{100.00}{2}
\emline{120.00}{125.00}{3}{120.92}{122.58}{4}
\emline{120.92}{122.58}{5}{121.75}{120.17}{6}
\emline{121.75}{120.17}{7}{122.48}{117.75}{8}
\emline{122.48}{117.75}{9}{123.12}{115.33}{10}
\emline{123.12}{115.33}{11}{123.67}{112.92}{12}
\emline{123.67}{112.92}{13}{124.12}{110.50}{14}
\emline{124.12}{110.50}{15}{124.48}{108.08}{16}
\emline{124.48}{108.08}{17}{124.74}{105.67}{18}
\emline{124.74}{105.67}{19}{124.92}{103.25}{20}
\emline{124.92}{103.25}{21}{124.99}{100.83}{22}
\emline{124.99}{100.83}{23}{124.98}{98.42}{24}
\emline{124.98}{98.42}{25}{124.87}{96.00}{26}
\emline{124.87}{96.00}{27}{124.67}{93.58}{28}
\emline{124.67}{93.58}{29}{124.38}{91.17}{30}
\emline{124.38}{91.17}{31}{123.99}{88.75}{32}
\emline{123.99}{88.75}{33}{123.51}{86.33}{34}
\emline{123.51}{86.33}{35}{122.93}{83.92}{36}
\emline{122.93}{83.92}{37}{122.26}{81.50}{38}
\emline{122.26}{81.50}{39}{121.50}{79.08}{40}
\emline{121.50}{79.08}{41}{120.00}{75.00}{42}
\emline{30.00}{75.00}{43}{29.08}{77.42}{44}
\emline{29.08}{77.42}{45}{28.25}{79.83}{46}
\emline{28.25}{79.83}{47}{27.52}{82.25}{48}
\emline{27.52}{82.25}{49}{26.88}{84.67}{50}
\emline{26.88}{84.67}{51}{26.33}{87.08}{52}
\emline{26.33}{87.08}{53}{25.88}{89.50}{54}
\emline{25.88}{89.50}{55}{25.52}{91.92}{56}
\emline{25.52}{91.92}{57}{25.26}{94.33}{58}
\emline{25.26}{94.33}{59}{25.08}{96.75}{60}
\emline{25.08}{96.75}{61}{25.01}{99.17}{62}
\emline{25.01}{99.17}{63}{25.02}{101.58}{64}
\emline{25.02}{101.58}{65}{25.13}{104.00}{66}
\emline{25.13}{104.00}{67}{25.33}{106.42}{68}
\emline{25.33}{106.42}{69}{25.62}{108.83}{70}
\emline{25.62}{108.83}{71}{26.01}{111.25}{72}
\emline{26.01}{111.25}{73}{26.49}{113.67}{74}
\emline{26.49}{113.67}{75}{27.07}{116.08}{76}
\emline{27.07}{116.08}{77}{27.74}{118.50}{78}
\emline{27.74}{118.50}{79}{28.50}{120.92}{80}
\emline{28.50}{120.92}{81}{30.00}{125.00}{82}
\put(17.00,57.33){\makebox(0,0)[lc]
{\small Fig.4: A scheme of conversion of an electromagnetic laser beam }}
\put(20.67,52.33){\makebox(0,0)[lc]
{\small exited in an open resonator into gravitational one and back; }}
\put(20.67,47.33){\makebox(0,0)[lc]
{\small $M1$, $M2$, $M3$, $M4$: Mirrors; $N$, $S$: magnetic poles; $BE$: Beam }}
\put(20.67,42.33){\makebox(0,0)[lc]
{\small Envelops of an electromagnetic and gravitational radiations.}}
\emline{25.67}{110.00}{83}{29.15}{109.32}{84}
\emline{29.15}{109.32}{85}{32.63}{108.69}{86}
\emline{32.63}{108.69}{87}{36.11}{108.11}{88}
\emline{36.11}{108.11}{89}{39.59}{107.58}{90}
\emline{39.59}{107.58}{91}{43.07}{107.10}{92}
\emline{43.07}{107.10}{93}{46.55}{106.66}{94}
\emline{46.55}{106.66}{95}{50.03}{106.28}{96}
\emline{50.03}{106.28}{97}{53.51}{105.95}{98}
\emline{53.51}{105.95}{99}{56.98}{105.67}{100}
\emline{56.98}{105.67}{101}{60.46}{105.43}{102}
\emline{60.46}{105.43}{103}{63.94}{105.25}{104}
\emline{63.94}{105.25}{105}{67.42}{105.12}{106}
\emline{67.42}{105.12}{107}{70.90}{105.03}{108}
\emline{70.90}{105.03}{109}{74.38}{105.00}{110}
\emline{74.38}{105.00}{111}{77.86}{105.02}{112}
\emline{77.86}{105.02}{113}{81.34}{105.08}{114}
\emline{81.34}{105.08}{115}{84.82}{105.20}{116}
\emline{84.82}{105.20}{117}{88.30}{105.36}{118}
\emline{88.30}{105.36}{119}{91.78}{105.58}{120}
\emline{91.78}{105.58}{121}{95.26}{105.84}{122}
\emline{95.26}{105.84}{123}{98.74}{106.16}{124}
\emline{98.74}{106.16}{125}{102.22}{106.52}{126}
\emline{102.22}{106.52}{127}{105.70}{106.94}{128}
\emline{105.70}{106.94}{129}{109.18}{107.40}{130}
\emline{109.18}{107.40}{131}{112.66}{107.91}{132}
\emline{112.66}{107.91}{133}{116.14}{108.48}{134}
\emline{116.14}{108.48}{135}{119.61}{109.09}{136}
\emline{119.61}{109.09}{137}{124.33}{110.00}{138}
\emline{124.33}{90.00}{139}{120.85}{90.68}{140}
\emline{120.85}{90.68}{141}{117.37}{91.31}{142}
\emline{117.37}{91.31}{143}{113.89}{91.89}{144}
\emline{113.89}{91.89}{145}{110.41}{92.42}{146}
\emline{110.41}{92.42}{147}{106.93}{92.90}{148}
\emline{106.93}{92.90}{149}{103.45}{93.34}{150}
\emline{103.45}{93.34}{151}{99.97}{93.72}{152}
\emline{99.97}{93.72}{153}{96.49}{94.05}{154}
\emline{96.49}{94.05}{155}{93.02}{94.33}{156}
\emline{93.02}{94.33}{157}{89.54}{94.57}{158}
\emline{89.54}{94.57}{159}{86.06}{94.75}{160}
\emline{86.06}{94.75}{161}{82.58}{94.88}{162}
\emline{82.58}{94.88}{163}{79.10}{94.97}{164}
\emline{79.10}{94.97}{165}{75.62}{95.00}{166}
\emline{75.62}{95.00}{167}{72.14}{94.98}{168}
\emline{72.14}{94.98}{169}{68.66}{94.92}{170}
\emline{68.66}{94.92}{171}{65.18}{94.80}{172}
\emline{65.18}{94.80}{173}{61.70}{94.64}{174}
\emline{61.70}{94.64}{175}{58.22}{94.42}{176}
\emline{58.22}{94.42}{177}{54.74}{94.16}{178}
\emline{54.74}{94.16}{179}{51.26}{93.84}{180}
\emline{51.26}{93.84}{181}{47.78}{93.48}{182}
\emline{47.78}{93.48}{183}{44.30}{93.06}{184}
\emline{44.30}{93.06}{185}{40.82}{92.60}{186}
\emline{40.82}{92.60}{187}{37.34}{92.09}{188}
\emline{37.34}{92.09}{189}{33.86}{91.52}{190}
\emline{33.86}{91.52}{191}{30.39}{90.91}{192}
\emline{30.39}{90.91}{193}{25.67}{90.00}{194}
\emline{50.00}{135.00}{195}{50.00}{125.67}{196}
\emline{50.00}{125.67}{197}{100.00}{125.67}{198}
\emline{100.00}{125.67}{199}{100.00}{134.67}{200}
\emline{100.00}{134.67}{201}{100.00}{134.67}{202}
\emline{100.00}{65.00}{203}{100.00}{75.00}{204}
\emline{100.00}{75.00}{205}{50.00}{75.00}{206}
\emline{50.00}{75.00}{207}{50.00}{65.00}{208}
\emline{50.00}{65.00}{209}{50.00}{65.00}{210}
\put(75.00,69.67){\makebox(0,0)[cc]{$S$}}
\put(75.00,130.00){\makebox(0,0)[cc]{$N$}}
\put(106.33,110.00){\makebox(0,0)[cc]{$BE$}}
\put(138.33,115.67){\vector(4,3){0.2}}
\emline{124.33}{110.00}{211}{126.69}{110.40}{212}
\emline{126.69}{110.40}{213}{128.99}{110.98}{214}
\emline{128.99}{110.98}{215}{131.23}{111.74}{216}
\emline{131.23}{111.74}{217}{133.39}{112.68}{218}
\emline{133.39}{112.68}{219}{135.50}{113.80}{220}
\emline{135.50}{113.80}{221}{138.33}{115.67}{222}
\put(138.33,84.33){\vector(4,-3){0.2}}
\emline{124.33}{90.00}{223}{127.57}{89.16}{224}
\emline{127.57}{89.16}{225}{130.42}{88.30}{226}
\emline{130.42}{88.30}{227}{132.88}{87.42}{228}
\emline{132.88}{87.42}{229}{134.96}{86.53}{230}
\emline{134.96}{86.53}{231}{136.63}{85.63}{232}
\emline{136.63}{85.63}{233}{138.33}{84.33}{234}
\put(138.33,97.00){\makebox(0,0)[cc]{$y$}}
\emline{74.67}{124.00}{235}{74.67}{76.67}{236}
\emline{74.67}{76.67}{237}{75.00}{76.67}{238}
\emline{75.00}{76.67}{239}{75.00}{124.00}{240}
\emline{75.00}{124.00}{241}{75.33}{124.00}{242}
\emline{75.33}{124.00}{243}{75.33}{76.67}{244}
\put(79.67,119.33){\makebox(0,0)[cc]{M3}}
\emline{114.67}{93.00}{245}{117.33}{90.33}{246}
\put(109.33,67.00){\vector(-1,-4){0.2}}
\emline{116.00}{91.67}{247}{109.33}{67.00}{248}
\put(120.00,130.67){\makebox(0,0)[cc]{M2}}
\put(30.00,130.67){\makebox(0,0)[cc]{M1}}
\put(118.67,87.00){\makebox(0,0)[cc]{$M4$}}
\end{picture}             \end{center}
             \vspace{-33mm}

The efficiency of the direct ways of the graviton production and
detection are less.

                  \section{Conclusion}
The direct way of emission of gravitational waves by charged ions
nonuniformly moving in external electromagnetic or gravitational fields
and conversion of the electromagnetic radiation into gravitational one
and back in the transverse magnetic fields have a different nature.
Both of them are in the Einstein's theory of gravity \cite{landau}.
Nevertheless second one is not so obvious. That is why they are need
personal verification in the emission and conversion processes. The
lack of the conversion process will not mean the lack of the
gravitational radiation emitted by the nonuniformly moving particles.

The power of any gravitational radiation source is very small and an
interaction of gravitational waves with any feasible detector is very
small. This is the reason why people were not able to confirm the
existence of such radiation. Nowadays the experimental efforts are
concentrated on the detection of Gravitational Radiation coming from
very intense astrophysical sources, such as Supernovae explosions, by
means of cryogenic resonant bars and interferometers. Artificial
sources of gravitational waves under human control will be the next
step in an accurate study the nature of the gravitational radiation,
its connection with another theories and its unification with other
forces. An experiment of laboratory production and detection of
gravitational waves would justify efforts comparable with those done
for particle accelerators. This happened in particle physics after the
introduction of accelerators beside of the use of cosmic ray detectors.

Unfortunately the intensity we can produce with the artificial
gravitational radiation sources proposed until now is much smaller
then requested by the sensitivity of the possible detectors actually
conceived. This could discourage on the possibility to realize an
experiment of production and detection of gravitational waves in
laboratory within not extremely long times. However some people are
optimist, and think that such an experiment will be done within the
next century. For this reason it is important to continue in
conceiving, studying and improving ideas for possible man made sources
of gravitational waves and for suitable detectors. If one rely on the
rapid technological increasing and on the continuous flow of new ideas,
it can hope the man will be able to do an experiments of such kinds
within next century \cite{chiarello}.

We hope that using of Grasers based on cold heavy ion beams (or more
heavy charged formations) and cutoff waveguides or on conversion of
electromagnetic waves in the transverse magnetic field as well as using
detectors with stimulation of graviton conversion by high quality
resonator and maybe another ideas (quantum nondemolition
measurements \cite{brag2}, using amplification by active media and so
on \cite{pisin1}, \cite{pisin2}) instill an additional hope in the
reality of the experiments on the verification of the gravitational
theories.

           \section*{Acknowledgments}

The author thanks Pisin Chen for invitation to discuss the problem at
the ICFA Workshop Quantum Aspects of Beam Physics QABP98 and Adrian
Melissinos and Kwang-Je Kim who paid attention at this Workshop on
papers \cite{pisin1}, \cite{pisin2}.

This work was supported by Russian Fund Fundamental Research, Grunt No
96-02-18167

\addcontentsline {toc} {section} {\protect\numberline {5
\hskip 2mm References}}

\end{document}